\documentclass[a4paper,11pt]{article}
\pdfoutput=1 

\usepackage{jcappub} 

\usepackage[T1]{fontenc} 

\usepackage{rcs} 
\usepackage{lineno}
\usepackage{graphicx}
\usepackage{amssymb}
\usepackage{amsmath}
\usepackage{times}
\usepackage{courier}
\usepackage{xspace}

\def\rmn{\ensuremath{\mathrm}}

\newcommand{\lat}{\emph{Fermi}-LAT}
\newcommand{\gr}{$\gamma$-ray\xspace}
\newcommand{\grs}{$\gamma$ rays\xspace}

\newcommand{\is}{interstellar}

\newcommand{\beq}{\begin{equation}}
\newcommand{\eeq}{\end{equation}}
\newcommand{\sig}{\ensuremath{\sigma}\xspace}
\newcommand{\sigv}{\ensuremath{\langle\sigma\nu\rangle_{\gamma\gamma}}\xspace}

\newcommand{\msol}{\ensuremath{M_{\odot}}} 
\renewcommand{\deg}{\ensuremath{^{\circ}}\xspace}
\newcommand{\gev}{\ensuremath{\mathrm{GeV}}\xspace}
\newcommand{\mev}{\ensuremath{\mathrm{MeV}}\xspace}
\newcommand{\mdm}{\ensuremath{m_{\mathrm{DM}}}\xspace}
\newcommand{\jointL}{\ensuremath{\mathcal{L}_{\mathrm{joint}}}\xspace}
\newcommand{\data}{\ensuremath{\mathcal{D}}}
\newcommand{\sigp}{\ensuremath{\boldsymbol{\mu}}\xspace}
\newcommand{\nup}{\ensuremath{\boldsymbol{\theta}}\xspace}
\newcommand{\smax}{\ensuremath{\mathrm{s}_{\mathrm{max}}}\xspace}

\title{Search for Gamma-ray Lines towards Galaxy Clusters with the \lat}
\author[a,b]{B. Anderson}
\author[a,b]{S. Zimmer}
\author[a,b,c]{J. Conrad}
\author[d]{M. Gustafsson}
\author[a,b]{M. S\'{a}nchez-Conde}
\author[e]{R. Caputo}
\affiliation[a]{The Oskar Klein Centre for Cosmoparticle Physics, AlbaNova, SE-106 91 Stockholm, Sweden}
\affiliation[b]{Department of Physics, Stockholm University, AlbaNova, SE-106 91 Stockholm, Sweden}
\affiliation[c]{Wallenberg Academy Fellow}
\affiliation[d]{Georg-August University G\"ottingen, Institute for theoretical Physics - Faculty of Physics, Friedrich-Hund-Platz 1, D-37077 G\"ottingen, Germany}
\affiliation[e]{Santa Cruz Institute for Particle Physics, Department of Physics and Department of Astronomy and Astrophysics, University of California at Santa Cruz, Santa Cruz, CA 95064, USA}
\emailAdd{brandon.anderson@fysik.su.se}
\emailAdd{stephan.zimmer@fysik.su.se}
\emailAdd{conrad@fysik.su.se}
\emailAdd{michael.gustafsson@theorie.physik.uni-goettingen.de}
\emailAdd{sanchezconde@fysik.su.se}
\emailAdd{rcaputo@ucsc.edu}
\abstract{
We report on a search for monochromatic \gr~features in the spectra of galaxy clusters observed by the \emph{Fermi} Large Area Telescope.  Galaxy clusters are the largest structures in the Universe that are bound by dark matter (DM), making them an important testing ground for possible self-interactions or decays of the DM particles. Monochromatic \gr lines provide a unique signature due to the absence of astrophysical backgrounds and are as such considered a smoking-gun signature for new physics. An unbinned joint likelihood analysis of the sixteen most promising clusters using five years of data at energies between 10 and 400 GeV revealed no significant features.  For the case of self-annihilation, we set upper limits on the monochromatic velocity-averaged interaction cross section. These limits are compatible with those obtained from observations of the Galactic Center, albeit weaker due to the larger distance to the studied clusters.
}

\begin{document}
\maketitle
\flushbottom


\section{Introduction}
Galaxy clusters (GCls) are the largest gravitationally bound structures in the Universe.  Their baryonic content, primarily in the form of a hot, dense, gas which emits thermal X-rays, is bound together by a much more massive halo of dark matter (DM).
It is typically found that 80\% of the total mass is comprised of DM, which is further supported by gravitational lensing analysis of GCls (see, e.g.,  \citep{Clowe:2006aa}). To date, DM has only been observed through its gravitational interactions (see e.g., \cite{Combes:2002aa,Bertone:2005aa,2009arXiv0901.0632E} for a review). Weakly interacting massive particles (WIMPs) make a good DM candidate largely because their thermal production in the early Universe naturally results in an abundance matching what we observe today (see, e.g. \cite{Bertone:2005aa,Feng:2010aa}).  The self-annihilation that guided this process could still carry on in regions of high DM density, producing standard-model particles including \grs (see, e.g. \cite{Bergstrom:2000aa,Bergstrom:2009ib} for a review). Such \grs~can be detected but, as baryonic matter falls into the potential well formed by DM, these dense DM structures provide in addition the host environment for interactions of cosmic rays with ambient gas which in turn can yield a sizable \gr~foreground.
One way to disentangle such potential foreground emissions is to focus on a subdominant, 
but unambiguous, WIMP process  --- the direct annihilation into monochromatic photons (see, e.g. \citep{Bergstrom:1988aa,Rudaz:1989aa,Giudice:1989aa}).

The flux ($\rmn{ph\,cm^{-2}\,s^{-1}}$) at Earth from a direct-to-photon annihilation is described by the formula,

\beq
\phi_{s}(\Delta\Omega)=\underbrace{\frac{1}{4\pi}\frac{\sigv}{m^{2}_\rmn{DM}}}_{\Phi_\rmn{PP}}\times\underbrace{\int_{\Delta\Omega}\int_{l.o.s.}\rho^{2} dl~{d\Omega}}_{\rm J-factor},\label{eq:dmflux}
\eeq

\noindent
where \sigv is the thermally-averaged cross section for DM annihilation into two photons, $m_\rmn{DM}$ is the WIMP mass,  $\rho$ is the DM density, and the integration is carried out over the solid angle and line of sight. A related process is the DM annihilation into $\gamma Z$ or $\gamma h$. In these cases one would expect to observe two lines, one located at the energy $E$ that corresponds to the mass of the WIMP and one at $E'=E\left(1-\frac{m^2}{4E^{2}}\right)$, where $m$ corresponds to the mass of the Z and Higgs bosons, respectively. \footnote{Note that we assume the WIMP to be a Majorana particle, which normally includes a factor $\frac{1}{2}$ in Eq.~\ref{eq:dmflux}, however, since we assume the WIMP to annihilate into two photons, this factor is cancelled out.}

It is convenient to separate the flux calculation of Eq.~\ref{eq:dmflux} into \emph{particle physics} ($\Phi_\rmn{PP}$) and \emph{astrophysical} (J-factor) components. Since the branching ratios DM annihilation into two photons are small however ($10^{-4}-10^{-1}$) \cite{Bergstrom:1997aa,Matsumoto:2005aa,Ferrer:2006aa,Gustafsson:2007aa,Profumo:2008aa}, in order to achieve a flux detectable by the \emph{Fermi} Large Area Telescope (LAT), the low-probability process must be compensated by a large J-factor. The Galactic center (GC) of the Milky Way is commonly expected to have the largest J-factor (see, e.g. \cite{Conrad:2015aa} for a recent review). Although the Galactic center of the Milky Way has the largest J-factor, GCls, though much more distant than the GC, have two advantages.  First, there are many of them, and the J-factor can be increased by analyzing them jointly, using a joint likelihood method \cite{Zimmer:2011aa,Huang:2012aa,Ando:2012aa,Griffin:2014aa}. The second stems from the fact that cold DM clumps together hierarchically, with large structures being comprised of small ones. These small DM clumps residing inside larger halos, also known as sub-halos, were formed earlier and are thus highly concentrated, which in turn means important contributions to the total DM annihilation flux from them. Indeed, because the annihilation rate depends on $\rho^{2}$, the J-factor is very sensitive to the detailed structure of these DM sub-halos. However, the DM (sub)halo hierarchy is only partially resolved by the cosmological N-body simulations used to infer DM density distributions (e.g. \cite{2008MNRAS.391.1685S,Kuhlen:2010aa}), and the relevant quantities must be extrapolated over many orders of magnitude in mass down to the smallest predicted substructures. When enclosing entire GCls in our regions of interest (ROIs), we observe the full effect of this extrapolation and the `boost' factor could potentially increase the expected line feature flux by factors into the thousands (e.g., \citep{Pinzke:2011aa}; see however \cite{MASC2013}).

Recent works have found hints for a line-like feature around 130~\gev in both the GC (see, e.g., \cite{Weniger:2012aa,Tempel:2012aa,Su:2012aa,Ackermann:2013aa}) and the most promising GCls (\cite{Hektor:2013aa}, hereafter referred to as HRT13). At such energies, the astrophysical foregrounds become negligible, making alternative (to new physics) explanations difficult. Nevertheless, the GC region has undergone particular scrutiny with detailed follow-up analyses suggesting the feature to be less significant than originally claimed \citep{Ackermann:2013aa} while the most recent re-analysis finds no supporting evidence for monochromatic lines from the GC \citep{Ackermann:2015aa}.

One of the last remaining undisputed supports of the DM interpretation for the $\sim$130 \gev feature is the claim by HRT13 that it is also present in the regions surrounding 18 of the most massive nearby GCls. There, a stacking analysis revealed hints for a double-peaked line at 130 and 110~\gev with a global significance of 3.6\sig. In light of the reduced significance of the GC excess \citep{Ackermann:2015aa}, we revisit the case of GCls, making use of additional data, an unbinned fit using both spatial and spectral information, and the technique of joint likelihood \citep{Anderson:2015aa,Conrad:2015aa}.

\section{Method}\label{sec:method}

\subsection{Data Selection}\label{subsec:data}
The LAT, main instrument aboard the \emph{Fermi} satellite, is a pair-conversion telescope sensitive to \grs in the energy range from 20~\mev\ to $>300~\gev$. For a more detailed description, the reader is referred to \cite{LAT}, and for the on-orbit performance to \cite{fermi_inst2012}. We analyze five years of public \lat\ Pass~7 reprocessed data taken between 2008-08-04 and 2013-08-04 in the energy range between 10 and 400~\gev. {Above 10~\gev the otherwise structured Galactic diffuse emission is less important and the LAT point spread function (PSF) is relatively narrow, with the on-axis 68\% containment radius being $<0.2\deg$ \citep{fermi_inst2012}. Above 400~\gev the number of detected photons is very low so we limit ourselves to lower energies.} In order to reduce the residual contamination of misidentified cosmic rays, we selected events passing the CLEAN-class selection cuts and use the associated P7REP\_CLEAN\_V15 instrument response functions (IRFs). We avoid \gr contamination generated by cosmic rays hitting the atmosphere of the Earth by removing events with a zenith angle $\theta_{z}>100\deg$ and excluding data taken during times when the field of view of the LAT came too close to the Earth limb (specifically we apply a cut on the LAT rocking angle $|\theta_{r}|\leq52\deg$). In addition we excise time periods of bright solar flares or \gr bursts and only select nominal science operations data. For the data preparation and analysis we use the publicly available {\tt{Fermi Science Tools}} version v9r34p0.\footnote{The data, associated software packages, and templates used to model the \is\ and extragalactic emission are publicly distributed through the \emph{Fermi} Science Support Center (FSSC) available at \url{http://fermi.gsfc.nasa.gov/ssc/data/}.}

Following earlier works \citep{Ackermann:2013aa}, we employ a sliding-window technique, in which we split the analysis into 128 energy windows. {Each fit occurs within a sliding energy window of $\pm 6\sigma$, typically corresponding to steps of half the LAT energy resolution $\sigma$ at the center of each window energy.} This window size is wide enough to contain an instrument response-convolved line, but narrow enough that the background can be simply approximated using a power law.

\subsection{Background Model}\label{subsec:bkgmodel}

Within our energy windows, the diffuse background can {be} described by a power law with free index $\Gamma$, and normalization, $n_{b}$ \cite{Weniger:2012aa,Ackermann:2013aa},

\beq
\phi_\rmn{iso}(\Delta\Omega,E)=n_{b}E^{-\Gamma}.
\eeq

\noindent
In order to improve convergence, we introduce the reasonable physical constraint $\Gamma \geq 0$. The diffuse background is a combination of emission from Galactic cosmic-ray interactions and unresolved extragalactic sources.  It is smoothly varying, and so for the small scales associated with our ROIs, we model it as an isotropic background.  The effects of these simplifying assumptions are discussed in \textsection{\ref{sec:systematics}}.

In addition to the diffuse {emission}, we model sources from the third \emph{Fermi} source catalog (3FGL, \cite{3FGL}) that are contained within our ROIs.\footnote{We model point sources up to $0.3\deg$ outside of our ROIs to account for the point spread function at 10~\gev.} We let the normalization $\boldsymbol{n}_{\rm{src}}$ of sources with detection significance greater than $10\sigma$ float, and fix those below.

\subsection{Signal Model}\label{subsec:sigmodel}
Since WIMPs are non-relativistic, direct annihilations into photons produce gamma rays that are monochromatic.  From there, we first account for the small cosmological redshifts to the GCls.  Second, we incorporate the LAT energy dispersion.
To do this, we simulate the redshifted lines (note that this effect is generally small since the farthest GCl is at $z<0.028$) according to the parametrized \lat\ IRFs, and use the output as our spectral model.  This method takes advantage of the standard \lat\ tools, in particular the likelihood fitting tool {\tt{gtlike}}, while also including the spatial information of an extended source.

Spatial GCl DM signal templates are derived by using the results from cosmological N-body DM simulations and linking those to the total mass determined from X-ray observables or gravitational lensing of each individual GCl (see, e.g. \cite{Kravtsov:2012aa} for a review).  In general, the distribution consists of a gravitationally binding host halo which is itself partly comprised of self-bound subhalos at a variety of mass scales. Given their formation times and accretion history, these subhalos are expected to be highly DM concentrated. As the annihilation rate is proportional to the square of the DM density, estimates of the total GCl flux are extremely sensitive to the level of this halo substructure. Incorporating local over-densities increases the total annihilation rate by a `boost' factor, $b$, compared with the prediction of the spherically smoothed DM distribution of the main halo. The determination of $b$ requires assumptions on both the relative abundance and concentration of all substructure masses (mass and concentration functions, respectively), see e.g. \citep{Lavalle:2008aa,Pinzke:2011aa,MASC2013}.
Although the DM density distribution of each individual sub-halo can be well-approximated by the Navarro-Frenk-White (NFW) \citep{Navarro1997} parametrization,

\beq \label{eq:nfw}
\rho(r)=\frac{\rho_{0}r_{s}^{3}}{r(r_{s}+r)^{2}},
\eeq

\noindent
with $r_s$ being the scale radius and $\rho_{0}$ the central density, there are uncertainties on the internal structure of sub-halos. We commonly introduce the concentration parameter $c_{200} \equiv r_{s}/R_{200}$ to describe the internal structure of DM halos, where $R_{200}$ is defined as the radius of the GCl where the enclosed density equals 200 times the critical density of the Universe. The uncertainty on this concentration parameter stems from the fact that N-body simulations can only resolve structure down to a few hundred times the size of their test particles, currently on the order of $10^{8}\msol$ for GCl-sized simulations \citep{Gao2012a,Hellwing:2015aa}.\footnote{High-resolution simulations can resolve these halo-mass scales at high redshifts \cite{Anderhalden:2013aa,Ishiyama:2014aa}, but the lack of GCl-size N-body simulation resolving the whole substructure hierarchy and the required extrapolations to present time still imply substantial uncertainties.} Depending on the details of the chosen model, substructures could exist at masses as low as $10^{-12}\msol$, meaning that the mass-concentration relation must be extrapolated over many orders of magnitude. The exact value of this cutoff is set by the kinetic-decoupling and baryonic-acoustic-oscillation damping, which ultimately depend on the particle physics nature of the DM candidate \cite{Green:2004aa,Profumo:2006aa,Bringmann:2009aa}. Here we adopt a common cutoff value of $10^{-6}\msol$ which has become a standard value in the field.

To account for the mass concentration uncertainty, we perform our analysis using two boost models --- generated by adopting bracketing extrapolations for the mass-concentration function. For the first we employ a power-law relation between DM-halo mass and concentration \citep{Gao2012a}.  This results in boost factors for individual GCls ranging from 240 to nearly 2000, and is hereafter referred to as our \emph{optimistic} configuration.
Alternatively, we follow the fiducial model from \cite{MASC2013} to obtain boost factors from 24--37 in our \emph{conservative} configuration.\footnote{We note though that power-law concentration models are strongly disfavored by recent developments in both the simulation side \cite{Anderhalden:2013aa,Ishiyama:2014aa,Hellwing:2015aa} and in our theoretical understanding of halo concentration at the smallest scales, e.g. \cite{MASC2013,Ludlow:2014aa}. Yet, we decided to include the {\it optimistic} boost in this work for a direct comparison with the results in HRT13. Note that the conservative model is not particularly sensitive to the precise choice of the substructure mass cutoff-value.} Compared to the pure NFW-case, which corresponds to a centrally peaked flux annihilation profile, the existence of substructure increases the expected \gr intensity towards the outskirts of the GCl. For the projected luminosity profile for annihilating DM, $I_\rmn{sub}$ from substructure as function of subtended angle $\theta \leq \theta_{200}$ from the center of the GCl, we adopt the functional form given by {Eq. 2} in \cite{Gao2012a}:
\beq \label{eq:substructure}
I_\rmn{sub}(\theta) = \frac{16b \times L_{\gamma}^{\mathrm{NFW}}}{\pi \ln{(17)}} \frac{1}{\theta_{200}^2 + 16\theta^{2}}.
\eeq
In the above equation we have introduced $\theta_{200}$ as the {angle substended by the GCl virial radius, i.e.\ } $\theta_{200}=\mathrm{arctan}(R_{200}/D_{a})\times180\deg/\pi$ where $D_{a}$ is the angular distance to the GCl. $L_{\gamma}^{\mathrm{NFW}}$ is the total \gr luminosity of the host halo, assuming an NFW profile (without substructure), in units of $\mathrm{ph\,s^{-1}\,sr^{-1}}$. We note that in this definition the non-boosted (NFW) scenario corresponds to $b=0$. Also, beyond $R_{200}$ we assume the predicted \gr signal to be negligible. We note that the fiducial model in \cite{MASC2013} does not provide a parametrization of the predicted luminosity profile. Yet, as it was shown in \cite{MASC2011}, even for moderate values of $b$, such as the ones proposed in \cite{MASC2013}, the flattening occurs in a similar way. We have compared the predicted profiles with those given by Eq.~\ref{eq:substructure} and find that they agree well. Hence we assume the same functional form for the different boost factor scenarios with only the value of $b$ varying. 

{\subsection{Target Selection}
  Starting from the extended HIFLUGCS \citep{Reiprich:2002,Chen:2007} catalog of X-ray flux-limited GCls, we first order potential targets by their J-factor (highest first), taking the first 16 GCls (note the discussion on the sample size in \textsection\ref{subsec:jointlike}). The location in the sky is show in Fig.~\ref{fig:gr_sky}.
\begin{figure}
\begin{center}
\includegraphics[width=0.999\columnwidth]{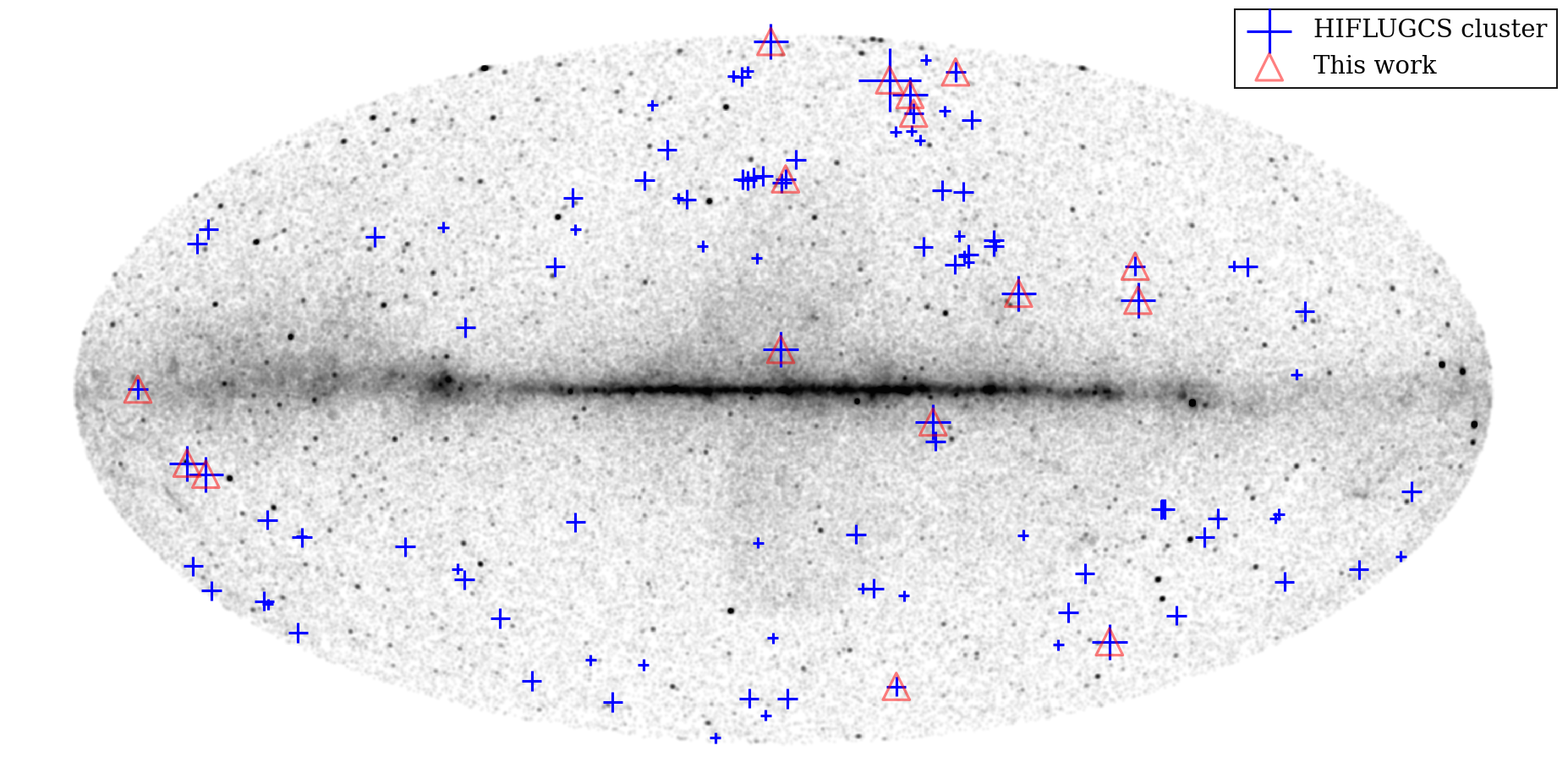}
\end{center}
\caption{Hammer-Aitoff projection (in Galactic coordinates) of the \emph{Fermi}-LAT counts map above $10\,\gev$ for 5 years of LAT exposure. Blue crosses represent the position of the 106 GCls that are contained in the extended HIFLUCGS catalog with the cross size proportional to $\log_{10}{M_{200}/D_{L}^{2}}$ of the GCl (larger crosses being more massive, nearby systems), where $D_{L}$ refers to the luminosity distance in units of Mpc. The colored red triangles indicate the targets used in this study.
\label{fig:gr_sky}}
\end{figure}
Since the predicted number of photons from astrophysical backgrounds is small, and we can localize the signal with its expected spatial information, the primary concern when determining the ROIs surrounding each GCl is to avoid overlap.  Performing a joint likelihood with overlapping signal regions can lead to over-estimation of significance \citep{Ackermann:2014aa,Anderson:2015aa}.  
There are infinite combinations of angular radii $R$ which yield a non-overlapping 
set of ROIs, so we must impose an additional constraint.  Because we only model 
the central target in each fit, expected DM emission from nearby GCls is considered 
background.  We therefore choose to require that our set of ROIs maximizes 
the ratio of total target emission to nearby GCl contamination.  Switching 
between boost scenarios changes the relative GCl DM emission levels and results 
in two unique sets of optimal ROI.
The respective values of $R$ for each ROI and each substructure scenario, $R_\rmn{opt}$ (optimistic) and $R_\rmn{cons}$ (conservative), along with other GCl parameters are summarized in Table~\ref{tab:targets}}.\footnote{In regards to the aforementioned free sources, note the following exception to this procedure concerns A2877: for the larger configuration, $R_\rmn{opt}$, we include four sources towards the edge of the ROI which have $TS>100$. Freeing all of these led to covergence issues which we mitigated by fixing the normalization of the weakest of these sources. Generically, the numbers of free sources is typically low: between 1--5 per ROI for both configurations.}

\begin{table}
  \scriptsize

\begin{tabular}{lrrrrrrrrrrrr}
    \hline
        {GCl}&{$\alpha_{2000}$}&{$\delta_{2000}$}&{$z$}&{$M_{200}$}&{$R_{200}$}&{$\theta_{200}$}&{$c_{200}$}&{${J_{\mathrm{NFW}}}$}&{$b_{\mathrm{cons}}$}&{$b_{\mathrm{opt}}$}&{$R_{\mathrm{cons}}$}&{$R_{\mathrm{opt}}$} \\
        {}&{($^{\circ}$)}&{($^{\circ}$)}&{}&{($10^{14}\,M_{\odot}$)}&{(Mpc)}&{($^{\circ}$)}&{}&{($\mathrm{GeV^{2}\,cm^{-5}}$)}&{}&{}&{($^{\circ}$)}&{($^{\circ}$)}\\
        \hline
 3C~129    &  72.29 &  45.01 & 0.021 &  5.90 & 1.78 & 1.12 & 5.030 & 16.19 & 34 &  922 & 6.0 & 5.8 \\
 A~1060    & 159.20 & -27.53 & 0.013 &  2.72 & 1.37 & 1.38 & 5.180 & 16.38 & 32 &  681 & 2.6 & 2.7 \\
 A~1367    & 176.10 &  19.84 & 0.022 &  8.13 & 1.98 & 1.19 & 5.000 & 16.25 & 34 & 1045 & 3.0 & 3.0 \\
 A~2877    &  17.45 & -45.90 & 0.025 &  7.54 & 1.93 & 1.02 & 5.000 & 16.11 & 34 & 1014 & 9.4 & 9.9 \\
 A~3526    & 192.20 & -41.31 & 0.011 &  3.72 & 1.52 & 1.80 & 5.110 & 16.62 & 33 &  770 & 4.0 & 2.1 \\
 A~3627    & 243.90 & -60.91 & 0.016 &  5.38 & 1.72 & 1.41 & 5.050 & 16.40 & 34 &  889 & 2.5 & 1.0 \\
 AWM7     &  43.63 &  41.59 & 0.017 &  5.38 & 1.72 & 1.33 & 5.050 & 16.35 & 34 &  889 & 1.0 & 1.3 \\
 Coma     & 195.00 &  27.98 & 0.023 & 10.92 & 2.18 & 1.25 & 4.970 & 16.30 & 35 & 1172 & 6.2 & 3.3 \\
 Fornax   &  54.63 & -35.45 & 0.005 &  1.39 & 1.10 & 2.84 & 5.350 & 17.02 & 30 &  525 & 5.2 & 3.1 \\
 M~49      & 187.40 &   8.00 & 0.003 &  0.72 & 0.88 & 3.79 & 5.580 & 17.27 & 28 &  405 & 1.5 & 1.6 \\
 NGC~4636  & 190.70 &   2.69 & 0.003 &  0.19 & 0.56 & 2.43 & 6.150 & 16.88 & 23 &  240 & 1.2 & 1.7 \\
 NGC~5813  & 225.30 &   1.70 & 0.007 &  0.46 & 0.76 & 1.40 & 5.750 & 16.40 & 26 &  340 & 0.5 & 0.1 \\
 Ophiuchus & 258.10 & -23.38 & 0.028 & 42.44 & 3.43 & 1.63 & 5.020 & 16.55 & 36 & 1990 & 5.0 & 3.1 \\
 Perseus  &  49.65 &  41.52 & 0.018 &  6.66 & 1.85 & 1.35 & 5.020 & 16.37 & 34 &  966 & 1.4 & 1.1 \\
 S~636     & 157.50 & -35.32 & 0.009 &  1.69 & 1.17 & 1.69 & 5.300 & 16.56 & 30 &  566 & 3.5 & 2.7 \\
 Virgo    & 187.70 &  12.34 & 0.004 &  5.60 & 1.70 & 6.28 & 4.210 & 17.41 & 34 & 1299 & 1.0 & 0.5 \\
        \hline
  \end{tabular}
  \caption{List of GCl parameters. The columns from left to right are: right ascension and declination in J2000 epoch, redshift, mass contained in virial radius $R_{200}$, angular radius, $\theta_{200}$, NFW halo concentration parameter, integrated $J$-factor for a smooth NFW halo (logarithm), boost factor for conservative and optimistic substructure models and the associated individually optimized radii for each ROI according to the prediction from the substructure model. We derive both $M_{200}$ and $R_{200}$ from the reported values for $M_{500}$ and $R_{500}$ from the HIFLUGCS catalog \citep{Reiprich:2002,Chen:2007} and assuming a value of the concentration parameter $c_{200}$, given by the model of \cite{MASC2013}. The mass of Virgo is taken from \cite{Fouque:2001aa}\label{tab:targets}.}
\end{table}

\subsection{Joint Likelihood}\label{subsec:jointlike}
Following the precedent set by the \emph{Fermi}-LAT Collaboration's searches for DM in dwarf galaxies (\cite{2011PhRvL.107x1302A}, \cite{2013arXiv1310.0828T}), we maximize our sensitivity by combining the information from multiple targets in a joint likelihood.

Beginning with each individual target, we compute the unbinned Poisson likelihood as
{
\beq\label{like}
\mathcal{L}_{t}(\sigp,\nup_t|\data_{t}) = \exp{\left({-\int\int \lambda~dE~dr}\right)} \prod_{i}  \lambda(\boldsymbol{r}_{i},E_{i}).
\eeq
}
\noindent
{In the above equation we introduced $\lambda$, the count-distribution function for all observed photons, as the product of $\sigp$ (our parameters of interest, $\sigp=\{\mdm,\langle \sigma v \rangle_{\gamma\gamma}\}$) and $\nup$ (the list of nuisance parameters $\nup_t=\{n_{b,t},\Gamma_t,\boldsymbol{n}_{\rmn{src},t}\})$. The integrals cover the ROI and energy window, respectively.} We label the normalizations of the background sources $\boldsymbol{n}_\rmn{src}$, diffuse background $n_{b}$, as well as the power-law index for the diffuse background Gamma, with an index $t$ to indicate that these quantities vary and are determined for each target individually. We calculate $\lambda$ by summing our model components (i.e. background source normalizations, power-law spectral index and normalization as well as the normalization of the DM target), each convolved with the exposure and instrument response functions.\footnote{Done with the tool {\tt{gtdiffrsp}}.}  We find the maximum likelihood $\mathcal{L}_{t}(\sigp,\hat{\nup}_t|\data_t)$ with respect to the nuisance background parameters and define the joint likelihood as the product over each individual target likelihood, $\mathcal{L}_{t}$:
\beq
\jointL(\sigp|\boldsymbol{\data})=\prod_{t} \mathcal{L}_{t}(\sigp,\hat{\nup}_t|\data_{t}).
\eeq
\noindent
Our measure of significance is the test statistic (TS),
\beq\label{eq:TS}
\mathrm{TS} = 2~\text{ln}\left(\frac{\jointL(\hat{\sigp}|\boldsymbol{\data})}{\jointL(\sigp_{0}|\boldsymbol{\data})}\right),
\eeq
\noindent
or twice the difference in log-likelihood between the best-fit and null ($\sigv=0$) hypotheses.  According to the asymptotic theorem of Chernoff \citep{Chernoff1952}, for fixed WIMP mass, the TS should be $\chi^{2}$-distributed with a single degree of freedom (we scan over a series of fixed values of $m_\rmn{DM}$ with \sigv~being the only free parameter), and we set our confidence intervals accordingly.
For global significance, the DM mass and the boost setup become free parameters, increasing the degrees of freedom and trials factor. We assess both coverage and global significance through Monte-Carlo (MC) experiments and outline these results in Section \ref{sec:systematics}.

\begin{figure}
\begin{center}
\includegraphics[width=0.9\columnwidth]{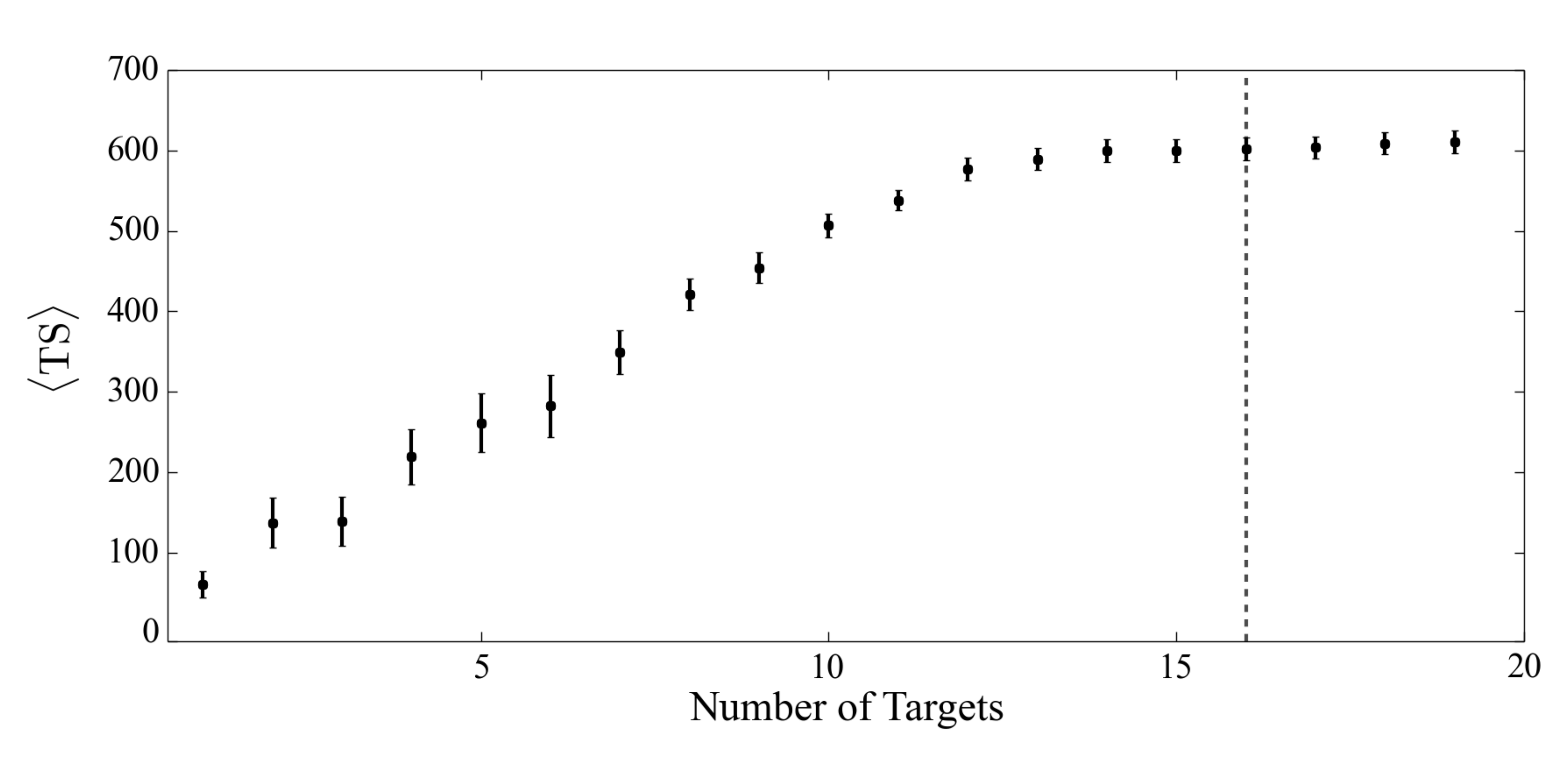}
\caption{\label{fig:tsvt} Evolution of the mean joint TS value derived from the combined analysis of GCls (in order of decreasing J-factor).  Based on ten simulations {with the conservative boost factor setup and} a $\sigv=1.1\times10^{-22}\mathrm{cm^{3}\,s^{-1}}$  for a 133~\gev monochromatic line with a background model comprised of large-scale diffuse emission templates and 3FGL catalog sources, respectively. The dashed line indicates the number of targets used in this analysis.  Injecting a signal with lower $\sigv$ causes the TS to plateau earlier.}
\end{center}
\end{figure}

\jointL could in principle be comprised of every possible GCl with no loss of sensitivity.  Targets from which we expect little to no flux are insensitive to changes in \sigp and cancel out of the TS.  However, to minimize both ROI overlap and computational burden, we form \jointL with a truncated set. We then compute TS as a function of the cumulative set for ten MC experiments that include a very strong ($\sigv=1.1\times10^{-22}\mathrm{cm^{3}\,s^{-1}}$) simulated line at 133 \gev.\footnote{Using the {\tt gtobssim} package.}  We place a cut at 16 targets (see Fig. \ref{fig:tsvt}), where more GCls would only contribute insignificantly to the TS.

\section{Performance and Systematic Uncertainties}\label{sec:systematics}

\subsection{Monte-Carlo Performance}
We set our significance and confidence interval for each value of \mdm under the assumption that the TS is distributed according to the asymptotic theorem around the maximum likelihood estimate of \sigv. A test of this assumption is warranted, in particular, given our simplified background model and the fact that many of our energy windows contain very few or even zero counts. To do this, we simulate a representative background which includes structured Galactic diffuse, isotropic, and point source emission.\footnote{To simulate the structured diffuse background we use the Galactic (gll\_iem\_v05\_rev1) and isotropic diffuse (iso\_clean\_v05) templates that are produced by the \emph{Fermi}-LAT collaboration and are distributed through the FSSC \url{http://fermi.gsfc.nasa.gov/ssc/data/access/lat/BackgroundModels.html}} By performing our analysis on a set of these simulations we calibrate our significance threshold. Then, following the spatial templates described in Section \ref{subsec:sigmodel}, we add DM lines resulting from a variety of \sigv to the simulation and assess our sensitivity and coverage.

\begin{figure}
\begin{center}
\includegraphics[width=\columnwidth]{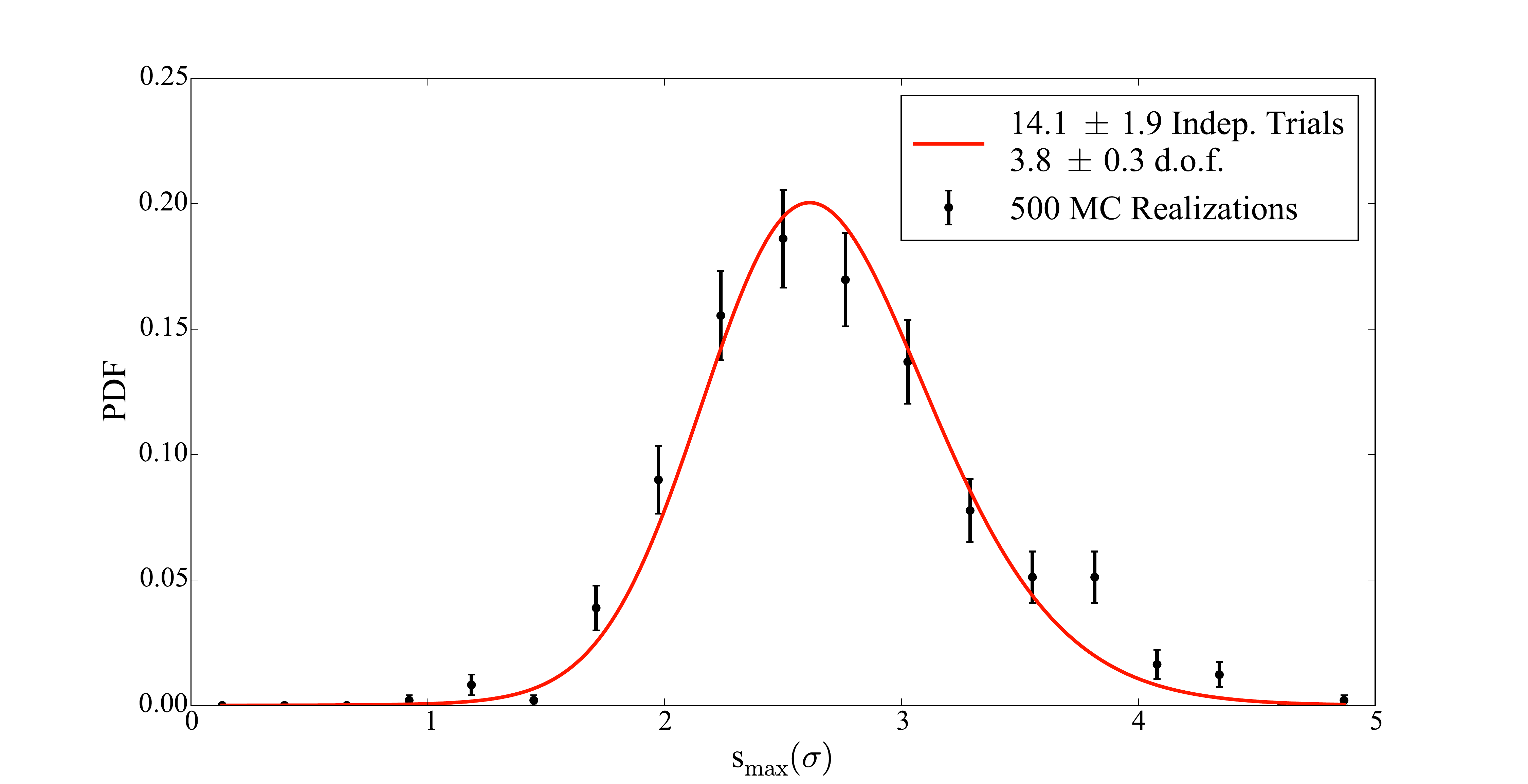}
\caption{\label{fig:trials}Probability distribution function (PDF) of maximum local significance ($s_\rmn{max}=\sqrt{\rmn{TS}}$) over all masses and boost scenarios for 500 MC experiments. Fit with a trial-corrected $\chi^{2}$ function, Eq.~\eqref{eq:trial_chi2}.}
\end{center}
\end{figure}

To determine the null distribution, and thus calibrate our significance, we perform a set of 500 background-only MC experiments for all 16 of our targets.  For each one, we extract the maximum significance ($\smax = \sqrt{\rm TS}$) from all energy windows and both \emph{optimistic} and \emph{conservative} boost scenarios. The resulting distribution is depicted in Figure \ref{fig:trials}.
To convert local to global significances we fit this with a trial-corrected $\chi^{2}$ distribution with both a free number of bounded degrees of freedom $k$, and trials $n_{t}$ \citep{Weniger:2012aa}:
\beq\label{eq:trial_chi2}
f(\smax) \equiv \frac{d}{dx}\rm CDF(\chi^{2}_{k};x)^{n_{t}}
\eeq

\noindent
CDF here means the cumulative distribution function.  The best-fitting parameters are $k=3.8\pm0.3$ and $n_{t}=14.1\pm1.9$.\footnote{Note that based on theoretical grounds, the expected number of degrees of freedom is two \citep{Gross:2010aa}.}

Using the calibration of global significance, we gauge our sensitivity to a weak DM signal, equivalent to the HRT13 claim.  To simulate this, we begin with a signal model which uses the spatial distribution and relative J-factor weighting of our \emph{conservative} setup.  {We then tune the value of $\sigv$ such that the total expected counts, including background, matches that reported in HRT13.}\footnote{A value of $\sigv=1.7\times10^{-24}\,\mathrm{cm^{3}\,s^{-1}}$ yields approximately 15 photons within a 5\deg radius for the stack of 18 HRT13 clusters with $125\,\gev\leq E\leq135\,\gev$.  Note that roughly half of these events are signal photons.}  We perform 10 MC experiments, keeping the total counts fixed, and find that we detect the feature with a local (global) significance of $4.0\pm0.6\sigma$ ({$2.4\pm0.7\sigma$}).\footnote{For the background component we use the same set of 500 MC simulations that were used to gauge the significance but \sigv was adjusted accordingly; however only a small fraction of the original number of MC simulations satisfies the selection criteria we use to map the analysis of HRT13.}

To confirm that our technique yields proper coverage, we further simulate lines for \sigv ranging from $3\times10^{-26}$ to $1\times10^{-21}$ cm$^{3}$ s$^{-1}$.  To make the calculation feasible, we perform 100 MC experiments for each line using only the four highest J-factor targets, and set joint upper limits.\footnote{These targets are Virgo, Ophiuchus, Fornax, and M49.	}
At high \sigv, we recover the expected 95\% containment.  Below a certain threshold in signal strength, however, we expect over-coverage as the sources predict no photons.  This occurs at approximately $3\times10^{-25}\mathrm{cm^{3}\,s^{-1}}$ (again for the conservative boost setup), and below that our method rises to 100\% coverage.  We conclude that our confidence intervals are conservative where they are not accurate.

\subsection{Systematics}
In this analysis we make use of the approximate representation of the LAT response via the IRFs. In order to assess how our results are affected by the uncertainties in both point spread function and effective area, we repeat our baseline calculation for the \emph{conservative} setup but use custom IRFs that bracket the associated uncertainties. These IRFs represent the minimal or maximal variations in the computation of the effective area and PSF within the systematic uncertainties of our chosen IRF (P7REP\_CLEAN\_V16). Specifically, these IRFs are chosen to maximize and minimize effective area and PSF, respectively (c.f.~\citep{fermi_inst2012} for details on the bracketing of PSF and effective area, respectively). Previous searches for \gr lines have shown that the uncertainties associated with the energy dispersion are much smaller than the statistical uncertainties of the analysis we report here \citep{Ackermann:2013aa,Ackermann:2015aa}. Hence we neglect this uncertainty in our bracketing IRF approach. The effect on our upper limits is minor, ranging between 10--20\%, with no obvious trend in energy. TS increases when all targets prefer the same value of the joint parameter, \sigv. Since the expected flux in Equation \ref{eq:dmflux} also depends degenerately on J-factor, TS is sensitive to this relative weighting of targets.  We explore our sensitivity to this fact by calculating \smax for both the HRT13-like signal simulations, and our LAT data, using identical J-factors for each target.  We find our local significance to be affected  on average by $0.2\pm0.2\sigma$, with a largest individual change of $-1.2\sigma$ (compared with the \emph{conservative} J-factor result) at $\mdm=52~\gev$.

A potential additional background source in each of the clusters is its brightest cluster galaxy (BCG). In some clusters, individual cluster member galaxies have been detected in \grs (e.g. NGC 1275 in Perseus \cite{Aleksic:2012aa} or M87 in Virgo \cite{Abdo:2009aa}). However, searches for \gr emission from a sample of 114 radioselected BCGs have only yielded null results \citep{Dutson:2013aa}. Even though the sample studied in \cite{Dutson:2013aa} differs from the one presented in this paper, we can use the observed average flux limit from \cite{Dutson:2013aa} to derive a conservative estimate of the BCG flux within our energy (and time) range. We find {$\overline{F}_{\gamma,\mathrm{BCG}}(E>10\gev)\lesssim2.0\times10^{-12}\,\mathrm{ph\,cm^{-2}\,s^{-1}}$}. This flux upper limit is comparable to the constraints on diffuse \gr emission $\overline{F}_{\gamma,\mathrm{ICM}}(E>10\gev)\lesssim2.7\times10^{-12}\,\mathrm{ph\,cm^{-2}\,s^{-1}}$ \citep{Ackermann:2014aa}, as it is expected from cosmic-ray interactions in the intracluster medium (ICM) of the galaxy clusters \citep{Pinzke:2011aa,Zandanel:2014aa}. The sum of these two contributions amounts to less than 0.1\% of the isotropic \gr background flux measured by the LAT in each ROI \citep{Ackermann:2015ab}. Converting these flux limits into photon counts, we find that the combined contribution corresponds to a total of $\sim20$ photons for the entire sample (with Virgo being the dominant cluster contributing $\sim8$ photons). Note that both scenarios (BCG and cosmic-ray induced \gr emission) assume a power-law like spectrum, in which case the dominant contribution would be expected towards the lowest energies and thus negligible compared to the observed number of photons in all ROIs (c.f. Fig.~\ref{fig:stacked_spectrum}).

\section{Results}\label{sec:results}
\begin{figure}
\begin{center}
\includegraphics[width=0.85\columnwidth]{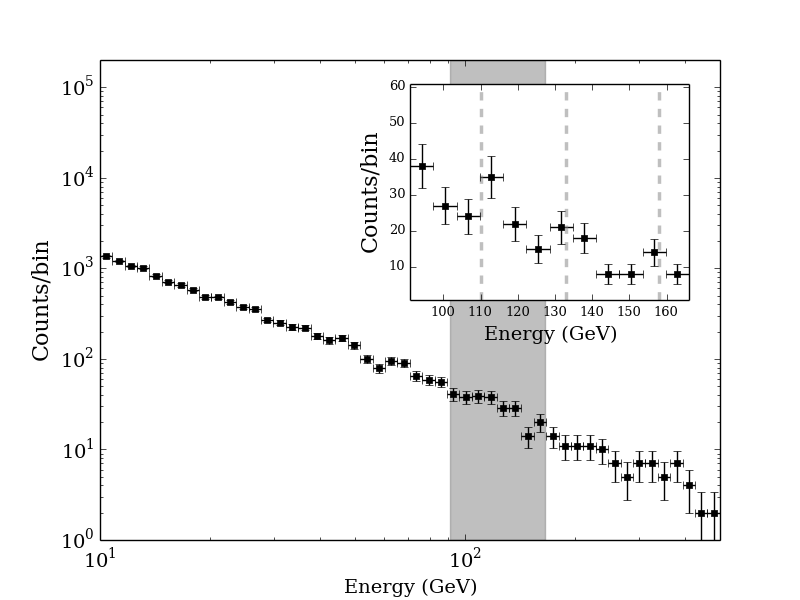}
\caption{\label{fig:stacked_spectrum} Shown is the stacked spectrum from photons sub-selected within each optimal radius of the 16 clusters analyzed in this work. The inset shows a zoom in the energy window (light-gray shaded area) that coincides with the discussed line signals at 110, 133, and 158 \gev, respectively (dashed-gray lines). Note that displaying the data as stack of binned histograms is for visualization purposes only, while the underlying analysis employs an unbinned joint likelihood method (see text).}
\end{center}
\end{figure}

\begin{figure*}
\hspace{-0.0cm}
\includegraphics[width=.52\textwidth]{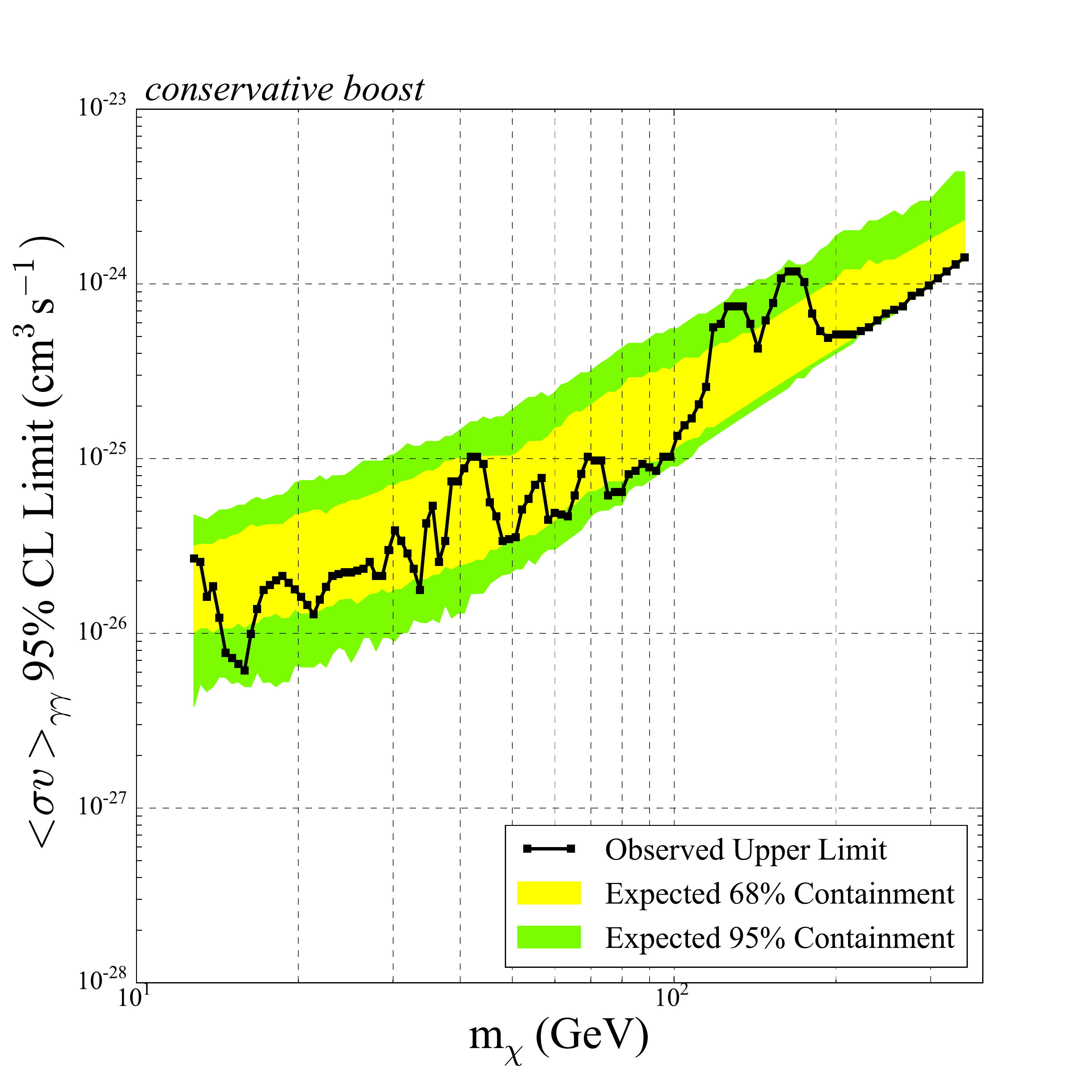}\hspace{-0.5cm}
\includegraphics[width=.52\textwidth]{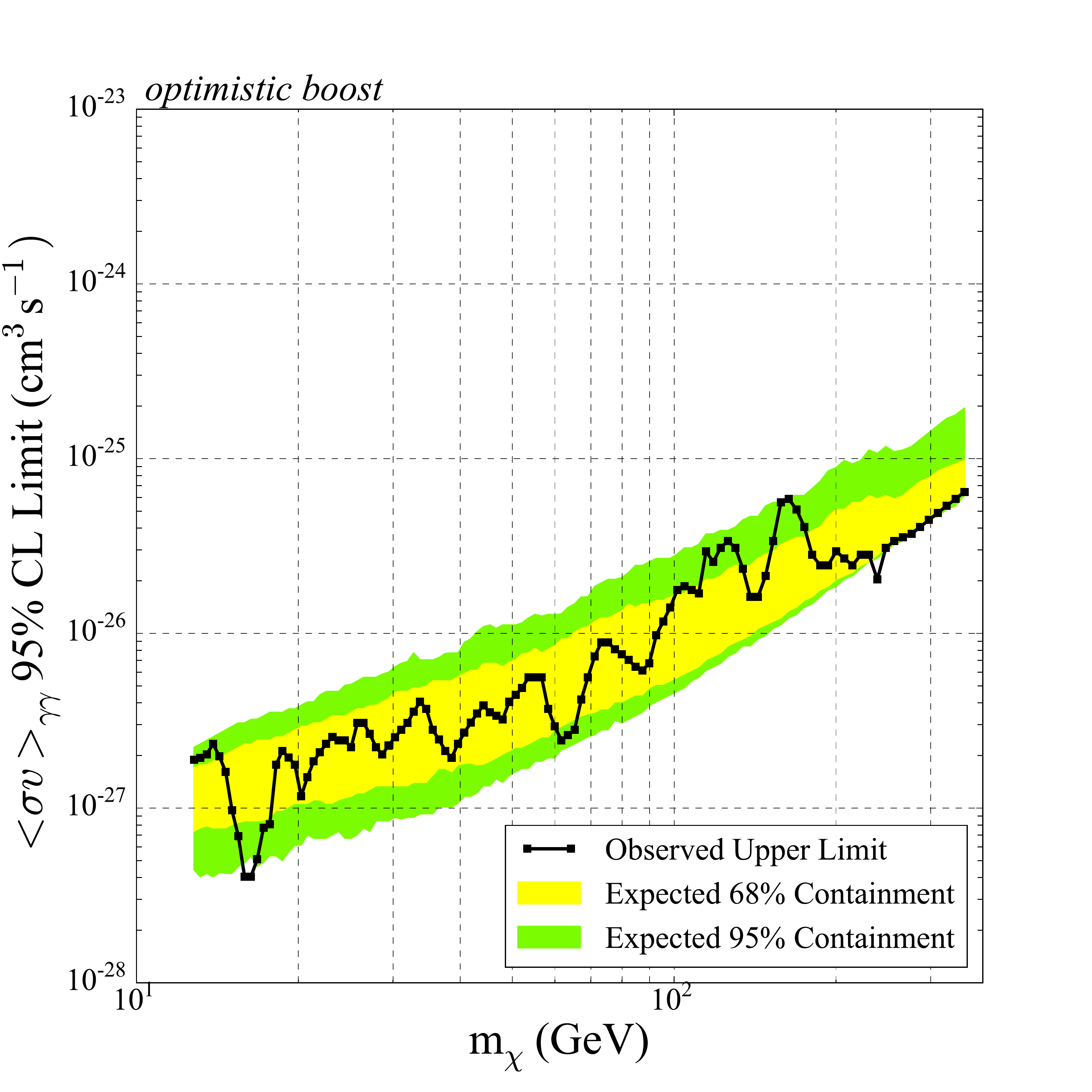}
\caption{Observed 95\% confidence upper limits for both \emph{conservative} (left) and \emph{optimistic} (right) boost scenarios.  Yellow and green bands represent 68\% and 95\% containment of limits obtained by MC simulations. Note that the upper limits and TS are not always correlated -- the upper limit value by itself gives no information about the relative likelihoods of the null and best-fit hypotheses.}
\label{fig:obs_limits}
\end{figure*}

Applying our analysis to a five-year dataset, we observe no significant excess over the energy range from 10 to 400 GeV and place 95\% confidence upper limits on \sigv.  The limits are plotted in Fig.~\ref{fig:obs_limits} for both boost scenarios, along with 68/95\% bands containing background-only MC limits. 

The stacked spectrum of all included photons can be seen in Fig. \ref{fig:stacked_spectrum}.  The highest local significance occurs for the \emph{optimistic} setup at 75 \gev with a value of $1.5\sigma$, corresponding globally to $<0.1\sigma$. These results are in agreement with recent updates on the GC \cite{Ackermann:2013aa,Ackermann:2015aa}, which also find no significant line emission.

It should be noted that the most recent of these GC studies \cite{Ackermann:2015aa}, utilized a new data set (Pass 8) which benefits from an improved instrument acceptance and PSF.  Performing our analysis with 6 years of P8\_SOURCE\_V6 data set yields results qualitatively compatible to those from Pass~7 reprocessed (this work), with a maximum local significance of $1.5\sigma$ at 181~\gev.  Determining the global significance would require repeating the studies of \textsection\ref{sec:systematics}, and would yield a lower value.  We also neglected to take advantage of a new feature, known as \emph{event types}, which subdivide the data based on the uncertainty of the energy reconstruction and provide corresponding IRFs.  Although this could have improved our sensitivity by approximately 15\% \citep{Ackermann:2015aa}, the lack of significant features in the data did not warrant its implementation.

\subsection[The Feature at $133\,\gev$ and Double Lines]{The Feature at $\mathbf{133\,\rmn{\mathbf{GeV}}}$ and Double Lines}
Although within the 95\% containment band expected from random background fluctuations, we note two upturns (around 133 GeV and 158 GeV) in both our observed limits between the energies of 100 and 200 \gev.
If the 133 GeV feature is indeed of the same origin as previously been seen in the GC \citep{Weniger:2012aa,Ackermann:2013aa}, where statistical fluctuations in Pass 7 data seem to have played a major role \cite{Ackermann:2015aa}, we expect to see a significant decrease in TS since the HRT13 claim.
Repeating our analysis with exposure periods of two, three and four years, respectively, we find the $\sim133\,\gev$ feature to decrease monotonically from $\mathrm{TS}\simeq1.8$ with 2 years of exposure to $\mathrm{TS}\simeq0.3$ with five years.

Inspired by the the HRT13 statements of a potential double peaked feature, from a $\sim$110 GeV and $\sim$130 GeV line, we also investigated if such a setup could increase the TS. With slightly wider energy windows,\footnote{We take the full energy range from the joint sliding windows from the two lines.} we repeat our analysis for the conservative boost scenario\footnote{Note that the optimistic boost factor scenario should give similar results.} with a second lower energy line together with a 133 GeV line.
We vary the relative intensity and take various energies of a second line around 110 GeV. We test 110 GeV as well as 106 GeV and 120 GeV, where the latter two energies are the expected energies if a 133 GeV line from DM annihilates into $\gamma\gamma$ is adjoined with a second line from $\gamma h$ or $\gamma Z$ final state, respectively. The TS for such double lines stays below $1$ in our joint-cluster analysis, and we conclude that with our analysis setup we see no significant single nor double line associated with a 133 GeV line. Similarly, we tested for a double line at 158 GeV and 133 GeV. These energies coincide with our observed two bumps in cross-section limits and can be theoretically expected if the DM particles annihilate into $\gamma\gamma$ and $\gamma h$. Such a double line can give a TS of 1.7, which is however still not statistically significant (especially if taking into account the ``look elsewhere'' penalty and that a larger energy window is used).F

\section{Discussion and Conclusions}\label{sec:discussion}

A detection of a monochromatic \gr~line from the most massive and nearby GCls would be extremely intriguing. The prime candidate for causing such a signal could only be the existence of DM particle annihilations or decays.

After a tentative line-like feature at 133 GeV has been reported in \emph{Fermi}-LAT data in the GC region \cite{Weniger:2012aa,Su:2012aa}, evidence was also presented that it was seen in GCls (HRT13).
In some DM particle models such strong line-like feature can indeed be expected in this energy region (see, e.g., \cite{Bergstrom:1988aa,Bergstrom:2009ib,Hisano:2003ec,Gustafsson:2007aa,Jackson:2009kg,Dudas:2012pb,Chu:2012qy,Buckley:2012ws}). Moreover, if a DM annihilation signal is confirmed towards GCls it would not only reveal important properties of the DM particle but also about how DM clusters  --- a detectable \gr~line signal in GCls can namely only be excepted if a large number of highly concentrated DM substructures exist down to small halo masses.

\begin{figure}
\begin{center}
\includegraphics[width=.8\textwidth]{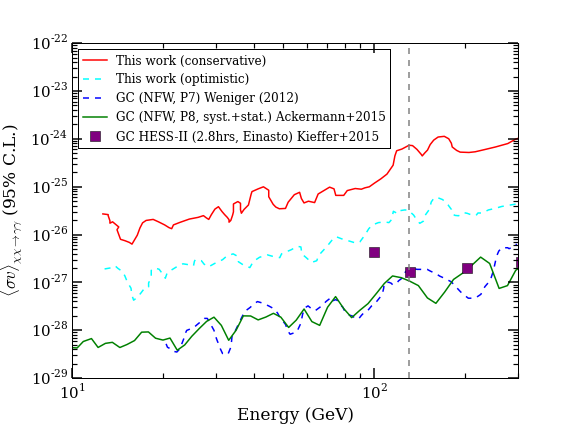}
\end{center}
\caption{\label{fig:otherlimits}Comparison of results presented in this work with other line searches; we only consider works in which constraints on $\sigv$ were calculated that overlap with the energy range used here. Green-solid and Blue-dashed lines correspond to results obtained analyzing data from the GC assuming an NFW profile using LAT data \citep{Weniger:2012aa,Ackermann:2015aa} while the squared markers correspond to recently published preliminary observed limits from 2.8 hours of observing the GC with the H.E.S.S. II telescope \citep{Kieffer:2015aa}. The gray-dashed vertical line corresponds to the energy at which the initial excess was reported $\sim130\,\gev$.}
\end{figure}

Subsequent \emph{Fermi}-LAT data have revealed that the statistical significance has dropped for the $\sim$133 GeV line signal from the GC \cite{Gustafsson:2013fca,Ackermann:2015aa}.
In this study we analyzed 5 years of {Pass~7 reprocessed data} to scrutinize the auxiliary support of the line feature from GCls. We find no remaining indications of a monochromatic \gr~line in our stacked cluster analysis. For different J-factor assumptions, we instead derive upper bounds on DM annihilation cross sections for DM masses in the range from 10 to 400 GeV. These limits (see Fig.~\ref{fig:otherlimits} for a comparison) are weaker than those {derived from the \emph{Galactic} \gr~line search program by the \emph{Fermi}-LAT collaboration} \cite{Abdo:2010nc,Ackermann:2012qk,Ackermann:2013aa,Ackermann:2015aa}, but they nevertheless constitute a very important complementary probe as long as the nature and clustering properties of the DM remain unknown.

\section*{Acknowledgments}
BA is supported by a VR excellence grant from the Swedish National Space Board (PI: Jan Conrad). MG acknowledges partial support from the European Union FP7 ITN Invisibles (Marie Curie Actions, PITN-GA-2011-289442). MSC is a Wenner-Gren Fellow and acknowledges the support of the Wenner-Gren Foundations to develop his research.

The \textit{Fermi} LAT Collaboration acknowledges generous ongoing support from a number of agencies and institutes that have supported both the development and the operation of the LAT as well as scientific data analysis. These include the National Aeronautics and Space Administration and the Department of Energy in the United States, the Commissariat \`a l'Energie Atomique and the Centre National de la Recherche Scientifique / Institut National de Physique Nucl\'eaire et de Physique des Particules in France, the Agenzia Spaziale Italiana and the Istituto Nazionale di Fisica Nucleare in Italy, the Ministry of Education, Culture, Sports, Science and Technology (MEXT), High Energy Accelerator Research Organization (KEK) and Japan Aerospace Exploration Agency (JAXA) in Japan, and the K.~A.~Wallenberg Foundation, the Swedish Research Council and the Swedish National Space Board in Sweden.
Additional support for science analysis during the operations phase is gratefully acknowledged from the Istituto Nazionale di Astrofisica in Italy and the Centre National d'\'Etudes Spatiales in France.

\bibliographystyle{JHEP}

\bibliography{jointline}
\end{document}